\newcounter{myctr}

\documentclass{ws-acs}
\usepackage{url}
\usepackage{color}
\usepackage[T2A]{fontenc}
\usepackage{soul}
\RequirePackage{color}

\DeclareMathAlphabet{\mathpzc}{OT1}{pzc}{m}{it}

 \definecolor{MyDarkGreen}{rgb}{0.02,0.60,0.06}

\newcommand{\be}{\begin{equation}}
\newcommand{\ee}{\end{equation}}
\newcommand{\bea}{\begin{eqnarray}}
\newcommand{\eea}{\end{eqnarray}}
\newcommand{\bem}{\begin{matrix}}
\newcommand{\eem}{\end{matrix}}

\begin{document}

\makeatletter
\def\@biblabel#1{[#1]}
\makeatother

\markboth{P. Sarkanych, N. Fedorak, Yu. Holovatch, P. MacCarron, J. Yose,  R. Kenna}{Network analysis of the Kyiv Bylyny cycle -- East Slavic epic narratives}

\catchline{}{}{}{}{}
%
%%%%%%%%%%%%%%%%%%%%%%%%%%%%%%%%%%%%%%%%%%%%%%%%%%%%%%%%%%%%%%%%%%%%

\title{NETWORK ANALYSIS OF THE KYIV  BYLYNY CYCLE \\ --  EAST SLAVIC EPIC NARRATIVES}

%\title{NETWORK ANALYSIS OF BYLYNY -- TRADITIONAL EAST SLAVIC EPIC NARRATIVES}

\author{PETRO SARKANYCH}

\address{Institute for Condensed Matter  Physics, National Acad. Sci. of Ukraine, Lviv, 79011, Ukraine\\
$\mathbb{L}^4$ Collaboration \& Doctoral College for the Statistical Physics of Complex Systems,
Leipzig-Lorraine-Lviv-Coventry, Europe
}

\author{NAZAR FEDORAK}

\address{Ivan Franko National University of Lviv, Lviv, 79000, Ukraine\\
Ukrainian Catholic University, Lviv, 79011, Ukraine
}

\author{YURIJ HOLOVATCH}

\address{Institute for Condensed Matter  Physics, National Acad. Sci. of Ukraine, Lviv, 79011, Ukraine\\
$\mathbb{L}^4$ Collaboration \& Doctoral College for the Statistical Physics of Complex Systems,
Leipzig-Lorraine-Lviv-Coventry, Europe\\
Centre for Fluid and Complex Systems, Coventry University, Coventry,
CV1 5FB, United Kingdom\\
Complexity Science Hub Vienna, 1080 Vienna, Austria
}

\author{P\'ADRAIG MACCARRON}

\address{MACSI, Department of Mathematics and Statistics, University of Limerick,\\
Limerick V94 T9PX, Ireland
}

\author{JOSEPH YOSE}

\address{Centre for Fluid and Complex Systems, Coventry University, Coventry,
CV1 5FB, United Kingdom\\
$\mathbb{L}^4$ Collaboration \& Doctoral College for the Statistical Physics of Complex Systems,
Leipzig-Lorraine-Lviv-Coventry, Europe
}

\author{RALPH KENNA}

\address{Centre for Fluid and Complex Systems, Coventry University, Coventry,
CV1 5FB, United Kingdom\\
$\mathbb{L}^4$ Collaboration \& Doctoral College for the Statistical Physics of Complex Systems,
Leipzig-Lorraine-Lviv-Coventry, Europe
}

\maketitle

\begin{history}
\received{(March 19, 2022)}
%\revised{(July 27, 2021)}
%\accepted{(Day Month Year)}
%\comby{(xxxxxxxxxx)}
\end{history}

\begin{abstract}
%The abstract should summarize the context, content
%and conclusions of the paper in less than 200 words.

In recent times, the advent of network science permitted new quantitative approaches 
to literary studies. 
Here we bring the Kyiv bylyny cycle into the field 
--- East Slavic epic narratives originating in modern-day Ukraine.
By comparing them to other prominent European epics, 
we identify universal and distinguishing properties 
of the social networks in bylyny. 
We analyse community structures and 
rank most important characters.
The method allows to bolster hypotheses from humanities 
literature --- such as the position of Prince Volodymyr 
--- and to generate new ones
We show how the Kyiv cycle of bylyny fits very well with
narrative networks from other nations — especially heroic ones.
We anticipate that, 
besides delivering new narratological insights, 
this study will aid future scholars and interested public to 
navigate their way through 
%one of Europe’s  major traditional epics  and identify its heroes.
 Ukraine's epic story and identify its heroes.

\end{abstract}


%%%%%%%%%%%%%%%%%%%%%%%%%%%%%%%%%%%%%%%%%%%%%%%%%%%%
\section{Introduction}\label{I}
%%%%%%%%%%%%%%%%%%%%%%%%%%%%%%%%%%%%%%%%%%%%%%%%%%%%

Over the past decade, 
quantitative methods have been developed to 
compare social networks depicted in  mythological narratives. 
In Refs.\cite{MacCarron2012} and other studies \cite{survey}, 
these have been applied to prominent European epics \footnote{For comparative purposes, by "epic" in this paper we  loosely mean an extensive narrative, that appears to describe events that happened far in the past, without demanding all scholastic hallmarks of epic (e.g. poetic form, formulaic composition, attested as an oral tradition, etc) are met. For the reader interested in authoritative explanations of the traditional oral epic, we refer to John Miles Foley’s Ref.\cite{JMF} and related literature. } such as 
in early Irish literature;
the Anglo-Saxon Beowulf; 
the ancient Greek Iliad; 
as well as a collection of sagas of the Icelanders \cite{MacCarron2013}. %,Kenna2016}. 
%In particular, it has been suggested to  treat the complex network of characters in a narrative in the way similar to that the 
%social network is treated. 
Here we bring bylyny --- heroic epics of eastern Slavs --- into the fold of network narratology. 
We consider epics from the heyday of Kyivan Rus' 
---  the end of the ninth century to the middle of the eleventh century. 
While narrative analysis is traditionally the prerogative of the humanities \cite{bod2013new}, 
and while societal-structure analysis is traditionally the subject of social sciences, 
the involvement of methods and concepts imported from the natural sciences is now broadly accepted in academia \cite{PNAS}.

Actually, applications of  quantitative methods to folklore and classical literature have a long tradition.
Historic-geographical approaches have been used to identify different types  of folk  tales, depending on topics, events and  characters~\cite{krohn1926}.
%\cite{aarne1961types}. 
The famous Aarne-Thompson-Uther catalogue includes more than two thousand such categories~\cite{atu}. 
A very different approach is the usage of principal-component analysis.
This allows 
%other interdisciplinary perspectives
%With this approach, 
each text to be presented in the form of a vector, with elements representing certain motifs ~\cite{Berezkin2017}.
Identification of similarity between texts is reduced to the measurement of similarities of vectors. 
%One of the interesting results obtained by this method is that the tales of the peoples of North America are different from the South American tales, which in turn may indicate that these continents were populated by people from the north and south directions independently \cite{Berezkin2017}. 
Another different, but equally innovative approach, is that of phylogenetics. 
This was developed for the study of evolutionary relationships between biological species 
%\cite{howe2011phylomemetics}
%In this approach, the inheritance of certain genetic traits is monitored 
%by the image of related species in the form of a graph. 
and adapted to search for connections 
%(the presence of common universal characteristics) in 
between various narratives~\cite{tehrani2013phylogeny,dhuy2015polyphemus}.
%\textcolor{blue}{Author in %Ref.~\cite{tangherlini2015} as well %provides network analysis of the Egils %saga, adding genealogical connections %to the table.
A popular quantitative approach is to  involve complex networks. 
E.g.  in  an  analysis of the Egils saga, Ref.~\cite{tangherlini2015}
 allowed sub-classifications of internodal relationships include geneaology, friendship, enmity interactions, gifts and inheritance. 
Another approach   
%Weingart an Jorgensen used network representation to 
analyses how adjectives are used with certain body parts in classic fairy tales \cite{weingart2011}. Network approaches can also be used to categorise tales \cite{abello2012} 
and a review of applications of networks to analyse fictional texts can be found in Ref. \cite{labatut2019}.

In  the  comparative approach used here, 
social networks are quantified by a set of 
observables such as node-degree distribution, 
clustering coefficients, assortativity, 
and other standard measures.
Each of these is not unique to social network, but can be used to  characteristic and compare them \cite{Newman2003}. 
%In turn, all these characteristics are widely used to create a basis on which comparison 
%and categorisation of social networks is conducted and to reveal the universal features 
%of social networks. In particular, 
By using these measures, it is by now well established that  social networks, 
in the main, have a number of common characteristics: they are mostly small-world
\cite{amaral2000classes,watts1998collective,milgram1967small,newman2001structure}; 
have heavy-tailed (right skewed) nodal degree 
distributions 
%are well described by power law
%are heavy tailed (i.e. right skewed)
\cite{amaral2000classes,barabasi1999emergence,newman2003random}; 
and their clustering coefficients are higher than would 
be expected for other (such as randomly constructed) network types~\cite{Newman2003}.
The social-balance hypothesis often holds for such networks too 
(``the enemy of my enemy is my friend'') \cite{heider1946attitudes,harary1953notion,Szell2010}; 
 they have hierarchical structures \cite{ravasz2003hierarchical};
and are assortatively mixed by node degree~\cite{Newman2003,newman2003structure}.
The program set up in \cite{MacCarron2012,MacCarron2013,MacCarron2014,Yose2017,Sarkanych2020} 
quantifies the extent to which networks of characters of mythological 
narratives that belong to different cultures exhibit these and other unifying and distinguishing features.
%Since all network characteristics can be 
%properly measured for any network of characters of any myth, they allow classification: 
%networks that share similar universal features are said to belong to the same (universality) 
%class. 
%In our study we follow the initiative of Refs. \cite{MacCarron2012,MacCarron2013,Kenna2016}, 
%first applying a similar method to 
Here we analyze bylyny - the heroic epic of Eastern Slavs. 
Continuing our studies initiated in Refs.~\cite{Sarkanych16,Sarkanych19}, our
goal is to find universal quantitative characteristics 
and to place bylyny in a larger context 
of mythological narratives of other cultures. 
We also wish to identify the most important characters and 
their placement  in the narrative, 
both amongst their immediate community and the broader society
--- ``the secrets to their success'', so to speak.
%We are interested in the 
%society described in the narrative, or more precisely, in the social network of characters. Thus, 
%the analysis of such a complex object as a text is reduced to the analysis of its representation 
%as a network. Moreover, since we are only analyzing one of the aspects of the narrative, our results 
%must be interpreted as complementary to those from traditional humanities approaches.
The rest of the paper is organized as follows. 
In  Section~\ref{II}, we briefly discuss bylyny, 
their content, as well as the period and place of their origins.
We present the complex network of bylyny society in Section~\ref{III} and rank most prominent  characters.
We discuss network properties in Section~\ref{IV}, where we compare
bylyny with network properties of other mythological narratives. 
We end with a conclusion and outlook in Section~\ref{V}.

%%%%%%%%%%%%%%%%%%%%%%%%%%%%%%%%%%%%%%%%%%%%%%%%%%%%%%%%%%%%%%
\section{Bylyny}\label{II}
%%%%%%%%%%%%%%%%%%%%%%%%%%%%%%%%%%%%%%%%%%%%%%%%%%%%%%%%%%%%%%

Bylyny  (or  staryny, as they are sometimes  
called)  are short, melodious, recitative epic songs of the Eastern Slavs. 
(The word ``bylyny'' is plural and its singular form is ``bylyna''.)
One of the first researchers of bylyny in the context of the Ukrainian medieval tradition was prominent historian and literary critic Mykhailo Hrushevsky who proposed the division of  bylyny into heroic-epic and religious-legendary parts \cite{Grushevskyi}. 
The first of these are older and are associated primarily with 
%the southwestern territories - primarily 
areas around Kyiv and Halych in modern-day Ukraine. 
They centre on  characters such as  Dobrynya, Illya Muromets', and Oleksiy (Oleshko, Alyosha) Popovych, as well as the so-called bylyny of the ``Halych-and-Volhynian group'' \cite{Grushevskyi}. 
These bylyny form the main focus of this paper.

The second group was created later are associated with Novgorod and Moscow.
These are social (not heroic) bylyny.
%are concentrated mainly around the elite.
%noble people, and they
Plots of the Novgorod cycle are grouped around ordinary people: merchants, moneylenders, musicians, etc.
%(harpists, guslari, etc.). 
Novgorod was an important (primarily naval) trade centre and, due to its geographical remoteness from land-based military and commercial routes, it evaded the Tatar-Mongol invasion in the 13th century. 
In contrast, Moscow was rapidly subordinated into the Golden Horde and  for several centuries the city was one of the western Mongolian ulus.
These historical and geographical details are reflected in bylyny local to different 
areas~\cite{Grushevskyi}.

%Bylyny can also be categorised according to the place where the action takes place. 

Given the thematic heterogeneity and striking differences in the geographic and political orientation of bylyny, it is not sensible to analyse the whole corpus as one here.
We  focus our attention only on the Kyiv cycle, which covers the period of the highest prosperity of Kyivan Rus’  (the end of the 9th to the the middle of the 11th century). 
Part of this epic tradition  includes the songs of skomorokhs (a typological analogue of Western European singer-minstrels)  and the story of  matchmaking. 
These epic works are not heroic, and therefore we also omit them from our analysis. 
The subject of our study is therefore 39 texts of epic tradition which we take from Ref. \cite{Putilov1986}.

%Message from Petro:
%In our draft we mention Kirsha Danilov to be the first one to publish bylyny. 
%He is known to be the first one, who collected around 70 bylyny and published them somewhere in the middle of 18th century. Before that, bylyny were passed down generation to generation orally. What is more, bylyny were sang, hence the oral spreading.
%Richard James also recorded some bylyny (or even just some fragments, don't remeber here tbh), but it wasn't a collection. They just appear in his diary.
%After the first publication of bylyny, they became rather a popular topic among scholars. The first issue they faced was "why bylyny tell story about Kyiv, but only a couple of them were recorded in the territory of modern Ukraine, while most of the texts were recorded in the northern russia hundreds of miles away from Kyiv". We touch upon this in the second chapter of the draft. And there are two answers to this question we mention.
%On the one hand, in 18th - early 19th century in Ukraine another type of epic songs, the dumas, took over that niche. Dumas originate from around 16th century and tell about cossacks, a phenomenon absent in russian culture. Therefore in Ukraine they were able to replace bylyny as something newer, but not in russia, as they were irrelevant for them. Interestingly, we have dumas about Illya Muromets', one of the most popular bylyny character.
%On the other hand, and this is what Avenarius said (we cite him), at the time Ukraine was literate an this limited information that was passed orally.

Bylyny were passed from generation to generation in oral form. 
Although the first recording was of the Moscow cycle  
by Englishman Richard James, for his diary, in 1619 \cite{Larin1959}, 
it is believed that Kirsha Danilov was the first 
to publish a collection of bylyny in the middle of 18th century \cite{Danilov1977}. \footnote{This collection shows signs of edits and corrections from the collector \cite{Oinas1961}.}
These describe events that happened in Kyiv, but they had been transmitted and migrated by oral tradition, and were collected and documented in the northern eastern Slavic lands in the 18th-19th centuries. 
Vasyl Avenarius explained this by the fact that, at that time, the majority of the population of Ukraine (south-western Eastern Slavic territory) was already literate, in contrast to the inhabitants of the remote villages of the empire.
In Ukraine, therefore, knowledge and testimony of  historical memory were passed in writing \cite{Avenarius1902}. 
%The second reason for the displacement of bylyny from the folklore environment was that 
Also, for Ukrainians at that time, more recent (originating in the 15th-18th centuries) recitative lyric epics about Cossacks (the so-called dumas) became popular
%\darkgreen{and these had} replaced the older bylyny epics
\cite{Shevchuk2003}. 
Also, the characters of bylyny in Ukraine ``migrated'' to other folklore genres, such as folk tales or folk songs.
Interestingly, there are dumas about Illya Muromets', one of the most popular of the bylyny characters.

Approximately 100 years after Danilov's collection, the topic of bylyny was brought to attention of scholars again. This was due to the fact, that 
Pavel Rybnikov discovered an active bylyny tradition in the Olonets region of northern Russia \cite{Rybnikov}. At first, this was received with a fair share of scepticism similar of that shown towards Ossian \cite{Yose2016}. However, subsequent expeditions verified its authenticity  \cite{Oinas1961}.\footnote{As it was pointed by Oinas, during the Soviet time folklore research was heavily affected by ideology and propaganda \cite{Oinas1975}.
His work on Kalevipoeg \cite{Finic}, the Estonian epic cycle, is a good example of  epic poetry in which the bylyny are considered part of the broader traditional genre.
}

Despite temporal, geographic and thematic differences, 
bylyny have a lot in common and different stories have common characters. 
Bylyny of the Kyiv cycle include a series of works related to the exploits of the most famous heroes: 
Illya Muromets', Dobrynya and Oleksiy Popovych. 
It is this feature of bylyny that allows us to interpret the characters of this epic tradition as a  social network. 
The main character of bylyny of the Kyivan cycle is Prince Volodymyr, who performs the classical function of ``epic king'' in the manner of many  medieval heroic epics (like Charlemagne in the so-called ``royal cycle'' of the French ``chansons de geste''). 
His wife is Opraksiya (Apraksiya, Eupraxia) and she never appears in the epics separately from her husband. 
Three main battle heroes are Illya Muromets', Dobrynya and Oleksiy Popovych. 
Each are vassals of Prince Volodymyr and are in military service to him. 
The main antagonists of the three heroes are the Soloviy The Robber, the Tatars and various bandits. 
In addition to concrete historical references, the words ``Tatar'' or ``Tatars'' may refer to an enemy in a more general sense.

A most interesting ``epic hero'' from a literary point of view is Illya,  an old experienced warrior who does not always blindly perform the orders of  Prince Volodymyr and sometimes transforms them or even resists them. 
In contrast, Oleksiy is young and hot-tempered. 
This pair are reminiscent of the characters of the 
 German epic ``Hildebrandslied'', 
 only there the old hero Hildebrand was the father of his young antagonist Gadubranda. 
Dobrynya is the guardian (perhaps an uncle) of Grand Prince Volodymyr-Vasyl  the Great, the baptist of Rus’, and in bylyny he is depicted as Volodymyr's nephew \cite{Grushevskyi}. 
According to some scholars,  Dobrynya may have been the main protagonist of the whole epic cycle in the period when it started to be created~\cite{Grushevskyi}. We give some more information about  the character(s) of Dobrynya in Appendix~A.

Bylyny feature the usual epic hyperbole. 
For example, characters may show superhuman abilities:
Illya, Dobrynya, or Oleksiy  is able to defeat several thousand troops alone; Soloviy The Robber is able to demolish houses with his piercing whistle, and so on (compare Cuchulain from T{\'{a}}in B{\'{o}} C{\'{u}}ailnge and the Ulster cycle in Ireland, for example).
An interesting and controversial issue is the link between  characters of bylyny and their likely historical prototypes. 

Historians, folklorists and literary scholars tried to 
get past distortion in the tales to a historical basis, 
especially for the Kyiv 
cycle~\cite{Kotliarevskii1862,Veselovskii1881,Potebnia,Grushevskyi}. 
The key character is Prince Volodymyr himself.
His mental, ideological and geographical connection with the Baptist of the Kyivan Rus’ Volodymyr-Vasyl the Great (958-1015) is indisputable. 
Historical sources are primarily medieval works such as 
 ``The Tale of Bygone Years''~\cite{cross1953}, 
 ``The Lives of SS. Borys and Hlib''~\cite{Abramovych1916}, as well as the life of Volodymyr the Great himself \cite{Tuptalo2005}.
 %Boris and Gleb are the sons of Volodymyr, who perished under the time of the fratricidal war for the throne of Kyiv between the descendants of the prince immediately after his death in 1015)
% which acquired its final canonical form, however, as late as in the seventeenth and eighteenth centuries \cite{Tuptalo2005}.
The life of Prince Volodymyr the Great gradually idealized in the memory of people and became an example of the Christian way of life in the context of the ancient Ukrainian chronicle tradition. 
Later rulers of Rus’ - mostly Volodymyr's descendants – were compared to his example.  
It can be assumed that the gradual transition %to folklore
from the oral to the written transfer of the most important events in the history of people - that is, the gradual transition from folklorization to fiction of history - 
began with the 
%appearance of the 
Kyiv chronicles.
This transition  became systematic after 988, when, through baptism, Rus’ was %annexed
adsorbed into 
%the "civilized" 
other European states and  cultures. 
Chronicle history knows many glorious princes, but after the death of the Baptist of Rus’, the name ``Volodymyr'' 
%has always imposed on his bearer of the royal environment 
bore a special 
royal
responsibility giving
%and gave 
rise to special expectations of his actions.
%activities. 
%Illustrative 
Examples of the ``worthy Volodymyrs'', whose images are detailed and vividly written in the chronicle tradition and in memory, are Volodymyr Monomakh (1053-1125, initially the Prince of Chernihiv and Pereyaslavl, later - the Grand Prince of Kyiv,  the great-grandson of Volodymyr the Great) and Volodymyr Vasylkovych (1249-1288, Prince of Volhyn', the great-grandson of the Grand Prince of Kyiv Mstyslav Izyaslavych, who himself was the great-grandson of Volodymyr Monomakh).

%Interestingly, f
Each of these two princes --- Volodymyr Monomakh and Volodymyr Vasylkovych --- 
is depicted only positively in the Ukrainian chronicle tradition. 
The first, besides becoming a positive character of the Tale of Bygone Years \cite{cross1953}, the Kyiv Chronicle (12th century)\cite{heinrich1977} and even the Halych-and-Volhynian Chronicle (13th century)\cite{vlasto1975hypatian}, 
%remained 
was the only Ukrainian prince-writer. 
His works (``Testament of Volodymyr Monomakh'', 
a fragment ``Letter to Prince Oleh'' and ``Prayer'') 
were placed in the Laurentian edition of the Tale of Bygone Years under the  annalistic year 1096 (the year according to the chronicles) \cite{cross1953}. 
The main human characteristics of Volodymyr Monomakh in  literary and national memory were justice, Christian zeal, enlightenment, wisdom, charity and he is also remembered for tireless and invincible military campaigns, political 
strength and statesmanly thinking, aimed at uniting the Rus’ lands and their rulers. 
As for Volodymyr Vasylkovych, the Halych-and-Volhynian chronicle presents an extremely touching portrait of this prince --- ``a philosopher never seen in the land of Rus’' and who suffered an incurable illness~\cite{Machnovec1989}. 
Volodymyr Vasylkovych did not have children, 
but took care of his people as a father.
He distributed much of his property to churches and monasteries, and was  the embodiment of wisdom, humility and mercy. 
According to the chronicles, when, four months after his death, his tomb was opened (at the command of his widow), 
Volodymyr Vasylkovych's  body was found uncorrupted: 
``the fragrance was from the grave, and the smell is similar to the expensive scents'' \cite{Machnovec1989}.

Thus, Volodymyr the Great,  Volodymyr Monomakh, and Volodymyr Vasylkovych 
together carry a whole complex of positive features.
These  were implicitly approved in the public consciousness and merged into an epic manifestation as the unified ``epic ruler'', Volodymyr. 
Still, local temporal and thematic 
%fragmentation 
differences in bylyny of different cycles 
suggest corresponding temporal, territorial and ideological personifications.
%the same fragmentation of "the unique Volodymyr" into his, so to speak, temporal, territorial and ideological personification.
%Prototypes could be, for example, 
Volodymyr the Great is associated with bylyny of the 10th-11th centuries;
Volodymyr Monomakh  with stories of the 12th-13th centuries; and Volodymyr Vasylkovych  for the stories  in the 13th-14th centuries 
(especially of Halych-and-Volhynian cycle). 
One of the schemes outlined below for the distribution of the epicenter of bylyny suggests this triple division of  Prince Volodymyr.
Whether Prince Volodymyr is part-amalgam of these historic characters or not is an interesting question, and one 
%Questions such as these have 
that has been addressed exhaustively using 
traditional philological and historical methods.
Complex networks cannot resolve such issues, of course. 
Still, very similar roles of ``superheroes'' are interesting to explore in network terms, not least to gain insight into how people in medieval times sought to present their heroic icons, real or imaginary~\cite{MacCarron2012,PNAS}.

%to their readers (and the roles national icons play today). 
%++++ At least, the result can somehow promote the image of Prince Volodymyr (or, for example, Dobrynya) on the imaginary coordinate line: either in the direction of historical-chronicle, or in the direction of the prevalence of folklore-archetype over folklore-epic.}}

%%%%%%%%%%%%%%%%%%%%%%%%%%%%%%%%%%%%%%%%%%%%%%%%%%
\section{The Kyiv Bylyny Network}\label{III}
%%%%%%%%%%%%%%%%%%%%%%%%%%%%%%%%%%%%%%%%%%%%%%%%%%

The extraction and analysis of networks of interacting characters from epic narratives 
is a recent  quantitative approach to comparative narratology; 
see Ref.\cite{survey} for an extensive survey up to 2019 and Ref.\cite{PNAS} for more recent developments.
% previous version cited \cite{Newman2010} and \cite{costa2011analyzing} 
In it, individual characters in a text are represented by  nodes
and interactions  between them are represented by  edges.
% previous cited  \cite{newman2011structure})  % \cite{Bornholdt2006}, 
In the approach introduced in Refs.\cite{MacCarron2012}, 
and in subsequent literature \cite{MacCarron2013,Kenna2016},
two characters are deemed linked if they know or interact with each other in some way.
Multiple meticulous  human readings are required to construct a reliable network 
--- e.g. to infer whether characters have met before and or whether they present together in a small group
or, indeed, if the interactions between them are hostile or not. 
The manual method of  Ref.\cite{MacCarron2012} 
and attempts at automation 
form two (of four) focal points for extraction and analysis of character networks in  Ref.\cite{survey}.
In Ref.~\cite{elson2010} the authors suggest a method to construct a network of characters automatically by allowing a computer to detect points in the text where two characters communicate directly. Despite this method showing good precision, it captures only one of our conditions for a friendship connection to be made --- namely when
two characters are in conversation. 
There are other methods to analyse the text with the help of modern data science approaches, like the one presented in Ref. \cite{karsdorp2012} where the ranked list of characters is obtained automatically using named entity recognition. Despite this method showing high accuracy, it does not allow to extract the interactions between the characters. Thus, 
it is not yet possible to extract the full, 
nuanced network using automated means and 
we use the manual approach here.
Therefore we use human reading to construct a friendly link between two characters if they talk to each other, from the text it is obvious that they have met before or they are present in a small group. In contrast, the hostile link is added if two characters either fight each other or are in a state of war.

\begin{figure}[!ht]
 \centerline{\includegraphics[width=9.0cm]{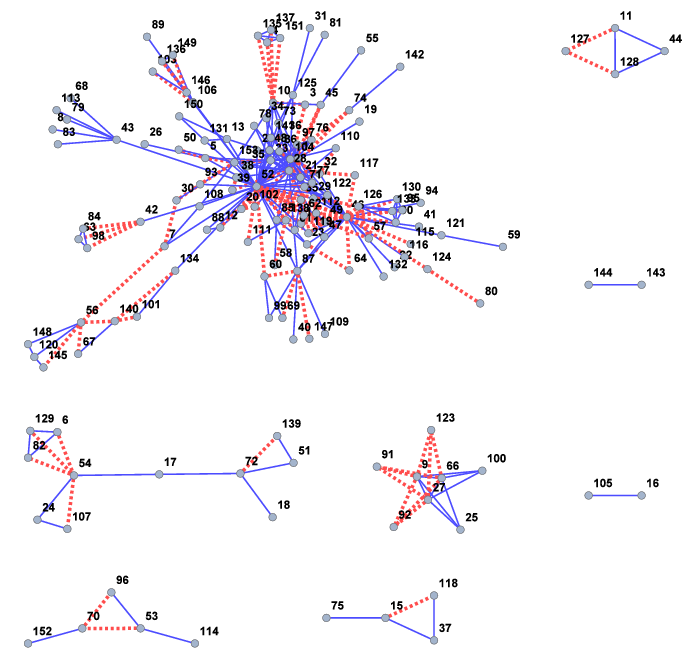}}
 \caption{Social network of characters of Kyiv cycle bylyny. 
 Red dashed lines corresponds to the hostile links, while blue solid to 
 the friendly ones. Each node has its own code number. The most important characters and the corresponding numbers are listed in Table~\ref{tab1}.
 The largest connected component is in the left top corner of the figure. Smaller graphs represent characters of separate stories. 
 For example, graph consisting of nodes 143 and 144 is two bulls observing troops who do not interact with other characters. The star shaped 
 graph in the second row from the bottom describes stories about Vavilo (node number 9). Characters of this part of the epics do not appear in any other story.}
 \label{fig1}
\end{figure}

From the 39 bylyny texts that form the basis of this study, we extracted a dataset comprising 153 individual characters connected by friendly and hostile links. 
This social network is depicted in Figure \ref{fig1}. 
Each character there is represented by a number and some of the most important ones in network terms are listed in Table~\ref{tab2}. 
The most prominent of these have already been mentioned and we encounter others in the sequel as well as in Appendix~A where we provide more background information to the Kyiv bylyny.  
Isolated nodes are omitted as they do not form part of a network. 
The terms used to rank characters are explained in the next section.
\begin{table*}[!b]
%\begin{ruledtabular}
    \centering
    {\small
    \begin{tabular}{|p{3.1cm}|p{0.4cm}|p{0.4cm}|p{0.4cm}|p{0.5cm}|}\hline
Character Name
& \#
&$R_k$
&$R_C$
&$R_B$ \\ \hline
Prince Volodymyr
&52
&1
&1
&1 \\
Illya Muromets'
&46
&2
&2
&2 \\
Dobrynya
&28
&3
&3
&3 \\
Oleksiy Popovych
&1
&5
&4
&10 \\
Mykhailo Potyk
&87
&9
&7
&5 \\
Princeess Opraksiya
&104
&4
&5
&13 \\
Duke
&33
&7
&6
&15 \\
Vasyl
&8
&10
&13
&7 \\
Churylo
&153
&6
&12
&16 \\
Soloviy the Robber
&126
&11
&15
&23 \\
Tatar
&138
&15
&14
&24 \\
Kudrevanko
&65
&10
&8
&37 \\
Dunai
&32
&18
&11
&31 \\
Ivan Hostynyi Syn
&43
&19
&37
&4 \\
Mykhailo
&85
&16
&25
&28 \\
Kalyn
&49
&17
&16
&36 \\
Matviy
&77
&12
&9
&49 \\
Luka
&71
&13
&10
&50  \\
Hoten
&146
&24
&40
&11 \\
Soloviy
&125
&25
&42
&14
 \\ \hline
            \end{tabular}}
    \caption{List of most important characters of bylyny, ranked by three commonly used network measures. 
    Character names are listed in the first column and their numeric identifiers (using the symbol \texttt{\#} in the text) are listed in the second.
    These numbers are sorted by the Cyrylic alphabet and used to locate the corresponding character in Figure~1. The three remaining columns give how each character is ranked in terms of node degree ($R_k$), closeness ($R_C$) and betweenness ($R_B$) centralities. 
}
    \label{tab2}
%\end{ruledtabular}
\end{table*}

In Figure~\ref{fig1}, friendly and hostile links are represented by solid blue lines and dotted red lines, respectively. 
Of the 320 edges, 223 are friendly and 105  hostile. 
The aggregate amount of friendly and hostile bonds is larger than the total number of edges because %each Bylyna itself is a short story 
%of several characters . In various 
 the same characters can interact in different ways
 (friendly and hostile) in different bylyny --- and even  within the same text \cite{MacCarron2013}.

%%%%%%%%%%%%%%%%%%%%%%%%%%%%%%%%%%%%%%%%%%%%%%%%%%%%%%%%%%%%%%%%%%%%%%%%%%%%%
%%%%%%%%%%%%%%%%%%%%%%%%%%%%%%%%%%%%%%%%%%%%%%%%%%%%%%%%%%%%%%%%%%%%%%%%%%%%%

The network in Figure~\ref{fig1} consists of several separate pieces.
The top left cluster is called the giant connected component. 
No individual bylyna has this many characters 
with some 10-20 characters in the biggest individual bylyna.  
{Similar ``subtle narrational tricks'' were used in 
modern narratives to mirror natural social networks~\cite{PNAS}.}
The size of the giant component  reflects the spread of 
characters across different tales.
This is not least because in the Kyiv epic cycle bylyny are united by the 
presence of hero-protagonist, Prince Volodymyr.
He is identified in the network as node number 52 
(we henceforth refer to node numbers using the symbol \#, so that Prince Volodymyr is identified as \#52).
In relative terms, the giant component comprises  76.5 \% of the bylyny network. 
This is smaller than for the epics of many other nations, but similar to that of the Old English epic poem Beowulf.
This comparison is not to suggest there is a connection between bylyny and Beowulf. 
Rather it is but one measure to position the complexity of bylyny relative to other epics. 
Our aim here is to quantify and document network features of bylyny and we do this using standard network indices --- precisely in order to compare to other epic narratives from other nations. 
The results are gathered in Table~\ref{tab1} where  giant components are represented by $G_c$ in the penultimate column. 

The other components of Figure~\ref{fig1} correspond to individual or smaller groups of stories.
These also belong to Kyiv cycle, 
but they don't share  characters with the largest connected component. 
For example the star shaped cluster in the second row from the bottom comes from the story about Vavilo (\#9). 
That tale relates how Vavilo overwhelms an antagonist king and takes his place 
(hence the high degree of hostility represented by dotted red edges). 
Characters in this story do not appear anywhere else, so they form isolated fragment of the full social network.
Thus, besides illustrating the cohesion and fragmentation of bylyny, Figure~\ref{fig1} also offers an aide to navigating one's way through the complex world it depicts.

The number of vertices is denoted by $N$ in Table~\ref{tab1} and the number of edges is $L$.
We explain the meaning of the other indices in Section~\ref{IV} where we use them for  comparative purposes.
The first row of the table contains statistics for the full social network of bylyny and results for the friendly and hostile networks are listed in the second and third rows. 
The hostile network has 98 nodes  
51\% (50) of which are in  the largest connected component. 
Bylyny have a higher proportion of hostile edges than the other narrative networks, data for which are also listed  in Table~\ref{tab1}.
Hostility for most epic tales tends to be  between 10\% and 20\% in terms of edge count.
The difference between bylyny and other tales can be explained by the fact that 
bylyny are 39 individual stories about heroes.
Each of the stories describes heroic exploits, they require an antagonist or even a couple of them for the hero to defeat. And with heroes being common for most of the stories, one may expect the portion of hostile links to be greater than in many other epics. 
%highlighting their difference in nature.

Different regions of the Kyivan Rus' had their own heroes, 
and these are not necessarily connected to the giant component. 
E.g., the  connected component  involving Vasyl' Okulovych (\#11), in a right upper corner of Figure \ref{fig1}, 
has a large proportion of hostility.

\begin{table*}[!b]
%\begin{ruledtabular}
    \centering
    {\small
    \begin{tabular}
        {|p{2.3cm}|p{0.4cm}|p{0.4cm}|p{0.4cm}|p{0.5cm}|p{0.4cm}|p{0.7cm}|p{0.4cm}|p{0.5cm}|p{0.6cm}|p{0.6cm}|p{0.7cm}|}\hline
    Network & $N$ & $L$ & $\langle k \rangle$ &
$k_{\rm max}$ & $l$ & $l_{\rm rand}$ & $l_{\rm max}$ & $C$ & $C_{\rm rand}$ & $G_c$ & $r_k$ \\ \hline
Bylyny (all)
& 153
& 320
& 4.18
& 52
& 2.9
& 3.61
& 6
& 0.57
& 0.03
& 76.5
& -0.15      \\
Bylyny (hostile)
&98
&102
&2.14
&17
&3.63
&5.76
&8
&0.06
&0.022
&51
&-0.11 \\
Bylyny (friendly)
&142
&223
&3.16
&41
&2.76
&4.33
&6
&0.42
&0.022
&57
&-0.11 \\
Beowulf (all) 
&72
&167
&4.45
&27
&2.38
&2.91
&6
&0.7
&0.06
&67.5
&-0.10 \\
Beowulf (hostile)
&31
&26
&1.67
&-
&2.08
&3.25
&4
&0
&0.05
&32.2
&-0.20\\
Beowulf (friendly)
&68
&141
&4.12
&-
&2.45
&2.98
&6
&0.69
&0.06
&66.1
&-0.03 \\
T{\'{a}}in  (all)
&422
&1200
&6.10
&168
&2.8
&3.3
&7
&0.8
&0.02
&98.5
&-0.33\\
T{\'{a}}in (hostile)
&144
&168
&2.33
&-
&2.93
&5.88
&7
&0.17
&0.02
&90.9
&-0.36\\
T{\'{a}}in  (friendly) 
&385
&1091
&5.67
&-
&2.84
&3.43
&7
&0.84
&0.01
&90.9
&-0.32\\
Iliad (all) 
&716
&2084
&7.40
&106
&3.5
&3.3
&11
&0.6
&0.01
&98.7
&-0.08\\
Iliad (hostile) 
&321
&361
&2.25
&-
&4.10
&7.12
&9
&0
&0.01
&89.4
&-0.39\\
Iliad (friendly) 
&664
&2317
&6.98
&-
&3.83
&3.34
&12
&0.62
&0.01
&82.3
&0.10\\
Odyssey
&301
&1019
&6.77
&112
&3.29
&3.18
&8
&0.45
&0.02
&98.3 
&-0.08\\
G\'isla Saga 
&103
&254
&4.9
&44
&3.4
&2.9
&11
&0.6
&0.05
&98
&-0.15\\
% Vatnsdaela Saga 
% &132
% &290
% &4.4
% &31
% &3.9
% &3.3
% &10
% &0.5
% &0.03
% &97
% &0.00\\
Egils Saga
&293
&769
&5.3
&59
&4.2
&3.4
&12
&0.6
&0.02
&97
&-0.07\\
Laxd{\ae}la Saga
&332
&894
&5.4
&45
&5.0
&3.5
&16
&0.5
&0.02
&99
&0.19\\
Nj\'als Saga
&575
&1612
&5.6
&83
&5.1
&3.7
&24
&0.4
&0.01
&100
&0.01\\
% Da Derga’s Hostel
% &120
% &410
% &6.51
% &71
% &2.70
% &2.77
% &7
% &0.04
% &0.05
% &98.4 
% &-0.18\\
% Nibelungenlied 
% &66
% &313
% &9.48
% &43
% &2.14
% &2.11
% &5
% &0.09
% &0.14
% &97 
% &-0.28\\
Mabinogion
&666
&2427
&7.29
&135
&3.83
&3.48
&11
&0.48
&0.01
&76
&0.19\\
% Tristan
% &49
% &133
% &5.43
% &44
% &1.99
% &2.40
% &4
% &0.75
% &0.11
% &100
% &-0.37\\
    % Epic of Gilgamesh
    % &40
    % &81
    % &3.52
    % &19
    % &2.54
    % &3.08
    % &5
    % &0.40
    % &0.08
    % &93.5
    % &-0.34\\
    % Popol Vuh
    % &98
    % &409
    % &8.35
    % &27
    % &2.80
    % &2.39
    % &0
    % &0.55
    % &0.09
    % &94.9
    % &-0.32\\
    % Navajo Myths
    % &140
    % &283
    % &4.04
    % &32
    % &3.81
    % &3.02
    % &9
    % &0.44
    % &0.03
    % &92.1
    % &-0.18
      \hline
            \end{tabular}}
    \caption{Characteristics of the social network of bylyny  
    compared to those of other epics (from Refs.~\cite{MacCarron2012,MacCarron2013,Kenna2016}).
    Here $N$ and $L$ indicate the number of characters and edges respectively.
    Other measures are 
    the average and maximum degrees ($\langle k \rangle$ and $k_{\rm max}$); 
    the average shortest path length of the complex network ($l$) and the counterpart for random graphs of the same size ($l_{\rm rand}$); 
   the diameter of the giant component  ($l_{\rm max}$); 
    the average clustering coefficient of the complex network and that of the corresponding random graphs ($C$ and $C_{\rm rand}$);  
   the size of the largest connected component  ($G_c$):
   and the assortativity  by degree ($r_k$). 
    Identifiers in brackets (``all'', ``hostile'', ``friendly'') specify types of relationships used to construct the network.
       }
    \label{tab1}
%\end{ruledtabular}
\end{table*}

For comparative purposes, 
Table~\ref{tab1} also contains network statistics for epics from other nations. 
Besides Ireland's T{\'{a}}in B{\'{o}} Cuailnge and  the Anglo Saxon Beowulf, mentioned previously, these include 
the Iliad and Oddysey from Greece and the Icelandic G\'isli Saga, %Vatnsdaela Saga,
Egils Saga, Nj\'als Saga, Laxd\ae la Saga and the Welsh Mabinogion.
% The Table also contains information on the characteristics of social networks of 
% Irish saga Da Derga's Hostel, Welsh Mabinogion, German Nibelungenlied, 
% Tristan and Isolde (Tristan), 
% Epic of Gilgamesh, 
% K'iche'Popol Vuh 
% and Navaho Myths.
The data for these texts come from previous publications~\cite{MacCarron2014}. 
When the information is available, we present data separately for friendly 
and hostile links as well as for the full networks. 
Comparison of the quantitative characteristics of social network of bylyny 
with those listed above enables a search for universal and differentiating 
characteristics of societal aspects of epic narratives.

In Ref.~\cite{MacCarron2014}, different categories of networks are
identified for different genres based on the shape of their degree
distributions and their assortativity. 
There it is observed that
stories centred on a single protagonist that travels to many
locations, such as Beowulf, the Odyssey or G\'isla saga all
are disassorative and the main characters having exceptionally
high degrees (relative to both system size and the rest of the
characters). 
On the other hand  narratives such as Laxd{\ae}la
saga, Nj\'als saga and the Mabinogion (which, after the first
four chapters mostly deals with Arthurian Romances) 
are set in one location, and while they have protagonists,
they are still  stories of a people. These have 
faster decaying degree distributions and are assortative. 
(We present a plot the cumulative degree distribution in Appendix~B.)
Between these extremes we find the Iliad and Egils saga, 
both of these have a few protagonists who are highly active
in the hostile network, these networks have assortative friendly networks. Similar to these is the T{\'{a}}in B{\'{o}} Cuailnge, however in this case we identify 6 characters with extremely large degrees in both hostile and freindly networks (for further discussion see Ref.~\cite{MacCarron2012}) placing it between these first and last categories. From the network measures in table~\ref{tab2}, we see that bylyny is more similar to that of Beowulf, the Odyssey and G\'isla saga. We discuss its individual properties and this result in more detail over the following sections.

%%%%%%%%%%%%%%%%%%%%%%%%%%%%%%%%%%%%%%%%%%%%%%%%%%%%%%%%%%%%%%%%%%%%%%%%%%%%%%%%%%%%%
\section{Network Statistics}\label{IV}
%%%%%%%%%%%%%%%%%%%%%%%%%%%%%%%%%%%%%%%%%%%%%%%%%%%%%%%%%%%%%%%%%%%%%%%%%%%%%%%%%%%%%

% % % % % % % % % % % % % % % % % % % % % % % % % % % % % % % % % % % % % % % % % % 
\subsection{Node degree}\label{IVa}
% % % % % % % % % % % % % % % % % % % % % % % % % % % % % % % % % % % % % % % % % % 

The degree of a node is the number of connections it has and, as such, 
it is one measure of a character's importance for the story. 
The node with the largest degree is Prince Volodymyr (\#52) who,
mentioned in almost every bylyna, 
is connected to roughly half of all other characters.  
At $k_{\rm max} = 52$, 
his degree  is much higher than the average degree $\langle k \rangle = 4.18$.
These maximal and average values are entered in Table~1 to give a feeling for how the central hero compares with central heroes of other texts.
These include Ireland's Cuchulain, England's Beowulf and Greece's {Achilles}.
While absolute values (of degree) are not meaningful comparative measures across different texts, rankings are, and Prince Volodymyr is highest ranked in Ukraine in the same way as Cuchulain is in Ireland and other national heroes are in their respective countries.
%, see Table \ref{tab1}. 
But Prince Volodymyr is not the only hero in Ukraine and 
%Although bylyny contain many characters, they are stories about heroes and the main attention is focused on these leaders. 
Illya Muromets' (\#46), Dobrynya  (\#28) and Oleksiy Popovych (\#1) are also prominent (see Appendix~A) and ranked accordingly. 
{In Appendix~A we present 
%\red{Figure~\ref{figA1}}
%The giant component in Figure~\ref{fig1} 
the giant component of Fig.~1 with nodes sized proportionate to node degrees.
Like Fig.~1, this is similar to a star graph, or rather to several star graphs fused  together; it draws out how leaders are directly linked to each other and also  attached to greater numbers of low degree characters.
(We also present a plot of the degree distribution in Appendix~B.)

As discussed in Section~2, some researchers believe Prince Volodymyr to be a collective character - a fusion of three distinct entities. We will return to this issue  in Subsection \ref{IVf} where we discuss the assortativity of the network \cite{Shevchuk2003,IEU}.

%\red{(For completeness, we also present a plot of degree distributions in Figure~X of Appendix~X.)}
%terms of degree.

% % % % % % % % % % % % % % % % % % % % % % % % % % % % % % % % % % % % % % % % % % 
\subsection{Distances between characters}\label{IVb}
% % % % % % % % % % % % % % % % % % % % % % % % % % % % % % % % % % % % % % % % % % 

Real social networks --- even those with large numbers of nodes --- 
are characterized by small average distances between them. 
This is the famous  small-world effect \cite{milgram1967small}. 
Since the 1960's the term ``six degrees of separation'' or ``six handshakes'' entered the public domain \cite{gurevitch1961social} and it is widely known that the average 
distance between randomly selected members of society is about six. 
More recently it was established that a  similar  value occurs in an equivalently sized Erd\H{o}s-R\'enyi random graph \cite{watts1998collective}.
Distance, in this sense, is the shortest length of a path or sequence of edges, which have to be traversed to get from one node to another. 
We denote it by $l$ and refer to such paths as geodesics. 
The longest geodesic is called the diameter of a network and represented by  $l_{\rm max}$.
We now check the typical distances (paths) between nodes in bylyny social network and compare to other stories from other countries.  
In the case of disconnected graphs like that of bylyny, we use only the giant connected component to determine diameters.

In order to find the average shortest path $l$ and diameter $l_{\rm max}$,
all geodesics in all connected components are taken into account
and the resulting values of $l$ and $l_{\rm max}$ are listed in Table \ref{tab1}. 
For bylyny the average path length is $l = 2.9$. 
Again this is most similar to the Irish T{\'{a}}in B{\'{o}} Cuailnge, for which $l = 2.8$.
The same can be said about the diameter of the network;
for bylyny  the maximum path length between characters is  6 
(for all and friendly networks) while for the T{\'{a}}in B{\'{o}} Cuailnge it is 7.
%In the real-world social network where the number of people-characters is of higher orders, the distances are the same because the real world is more connected. 
The value of the diameter $l_{\rm max}=8$ for the hostile network of bylyny is higher, 
which means that this network is `stretched' relative to the friendly one. 
A similar feature is  observed for other epics in Table \ref{tab1}.

It is usual to compare the average geodesic length $l$ 
to the corresponding value $l_{\rm rand}$ calculated 
for an Erd\H{o}s-R\'enyi random graph \cite{erdos1960evolution} of the same size. 
An Erd\H{o}s-R\'enyi graph is an example of the small
world network for which average path length grows 
logarithmically with the number of nodes \cite{bollobas1981diameter}. 
%This, and the  lack of correlation in node location, leads to low values of average path length.
Average values of $l_{\rm rand}$ are given in 
Table \ref{tab1} for random networks of appropriate sizes.
As one can see from the table, 
the average shortest distance for bylyny networks are lower than for 
random graphs with relevant number of edges and nodes.
This is similar to the Irish T{\'{a}}in B{\'{o}} Cuailnge
and  
%other Irish tale such as  Da Derga’s Hostel.
the Angle Saxon Beowulf 
%and Tristan have the same feature, as does Epic of
%Gilgamesh. 
Interestingly, however, the Icelandic sagas and Greek classics, 
%Nibelungenlied and the tales from the  K'iche' and Navajo people %(is it Navaho?) 
do not share this quality.

% % % % % % % % % % % % % % % % % % % % % % % % % % % % % % % % % % % % % % % % % % 
\subsection{Closeness centrality}\label{IVc}
% % % % % % % % % % % % % % % % % % % % % % % % % % % % % % % % % % % % % % % % % % 
 
We are also interested in the concept of distance between  characters.
One relevant measure is the closeness centrality \cite{Newman2010,brandes2001faster}. 
For each node, this is 
%in the graph, the closeness centrality is 
the inverse of the sum of the distances 
%from  the given node 
to all the other nodes within the connected component.
Therefore, the higher the closeness centrality $C_C$ is, 
the closer the given node is to all other nodes. 
%For meaningful comparisons, calculations should be performed only within the connected component of a network where there is a path between any two characters.
%In Panel (b) of Figure \ref{fig2}, we depict the largest connected component of bylyny social network. 
%Here the size of each vertex is proportional its $C_C$ value. 
See Appendix~{{B}} for a visual depiction of the giant component weighted according to this measure.
The top five characters ranked according to  closeness centrality are 
Prince Volodymyr (\#52);
Illya Muromets' (\#46); 
Dobrynya (\#28);
Oleksiy Popovich (\#1) 
and Princess Opraksiya (\#104). 

Immediately after these are Duke (\#33) 
and Mykhailo Potyk (\#87). 
``Duke'' here is a name and not a title.
The famous Ukrainian historian  Mykhailo Hrushevsky,
who 
%was a specialist 
specialised 
in the history of literature,
%referred to these two 
identified them 
as  part of the 
Halicia–Volhynia character group \cite{Grushevskyi}. 
According to bylyny, Duke is a Halician prince
and Mykhailo Potyk is considered to be borrowed from Western Ukraine, 
where he, in turn, came from Bulgaria, with a history of St. Michael 
from Potuka \cite{Grushevskyi}. 
{In the texts, %we took into consideration that - we tried to say "texts we read", thus commenting it out should do
he is a hero fighting 
on the side of Prince Volodymyr and against King  Lyahetskiy (\#60).} 
%According to another version, the name came from the river Potok in Western Ukraine,  from which, in turn, the Potocki dynasty took the its (wiki Potyk). this info is nowhere to be found - the citation we used earlier was to online source and is unavailable now
We find it interesting that closeness centrality has identified 
the importance of characters of the Halych-and-Volhynian region, 
which once was a successor of the Kyivan Rus'.
These characters have relatively low degrees.
It is also interesting that, without
Volodymyr and Opraksiya, the highest ranked character by closeness centrality is 
Dobrynya who outranks Illya Muromets'.
This may be a reflection of the fact that  in several texts Illya opposes Prince Volodymyr, while 
Dobrynya is always on the side of the ruler, and, therefore, closer to him.

Thus, besides a simple ranking of network nodes, 
individual metrics may inspire us to look deeper 
into individual characters, questioning the roles
they play in the story and how they are represented.
We continue in this vein with other measures of centrality.

% % % % % % % % % % % % % % % % % % % % % % % % % % % % % % % % % % % % % % % % % % 
\subsection{Betweenness centrality}\label{IVd}
% % % % % % % % % % % % % % % % % % % % % % % % % % % % % % % % % % % % % % % % % % 

Another local characteristic that can give an insight onto the 
importance of individual nodes is the betweenness centrality $C_B$. 
This measures how important the node is in maintaining communications between other 
 nodes. 
Betweenness centrality is the fraction of geodesics passing through a given node. 
Like the closeness centrality, betweenness makes sense only within  connected components.  
Accordingly, a value  $C_B = 1$ corresponds to the case when all the geodesics of the network  pass through the given node.
%In Panel (c) of Figure\ref{fig2} (Appendix X), nodal sizes are proportional to betweenness  centrality. 
The lead characters according to in betweenness terms are again Prince Volodymyr (\#52), Illya Muromets' (\#46) and Dobrynya (\#28). 
Oleksiy  Popovych (\#1) is ranked significantly lower according to this measure.
Interestingly, Ivan Hostynyi Syn (\#43) is ranked fourth although he appears just once, in a story which only has five unique characters. 
However, because Ivan has a direct connection to Prince Volodymyr (Figure~\ref{fig1} ) these five unique characters are connected to the rest of the network.

In ways like this, nodes with high betweenness centrality 
are important for maintaining the integrity of the network. 
For example, if one removes the three characters of 
bylyny social network with the highest values of $C_B$, 
the network splits into 24 pieces,  
the largest of which contains only 72 nodes. 
%Resistance of the network to removal of individual components 
%will be analysed in section \ref{IVh} of Appendix X.

% % % % % % % % % % % % % % % % % % % % % % % % % % % % % % % % % % % % % % % % % % 
\subsection{ Clustering coefficient}\label{IVe}
% % % % % % % % % % % % % % % % % % % % % % % % % % % % % % % % % % % % % % % % % % 

The clustering coefficient 
measures the probability that 
neighbours of a particular node 
in a network are themselves  connected  \cite{watts1998collective}. 
In a fully connected graph, the clustering coefficient of a node is equal to 1.
In a tree graph (one without loops) it is 0.
%As one can see from from 
%Table 1, the importance of a node ranked according to their clustering coefficients is significantly different from 
%their importance according to the degree.\footnote{\red{Seems like this column is missing from Table 1?} \blue{Can we just remove this sentence? }}
%, cf. Figure \ref{fig2}(a) and Figure \ref{fig2}(d). 
For bylyny, and in complex networks in general~\cite{ravasz2003hierarchical},  nodes with high degree tend to have small clustering coefficients; 
the larger the degree of the node is, the less likely that its neighbours are connected to each other. 
This is due to the fact that heroes [Illya (\texttt{\#}46),  Dobrynya (\texttt{\#}28), Oleksiy Popovych (\texttt{\#}1)] each have several different  enemies independent of each other.

While clustering coefficient is a local (nodal) characteristic,  
its average value tells us something about the nature of the whole network. 
In contrast to the manner in which complex networks tend 
to have similar average path lengths to that of Erd\H{o}s-R\'enyi graph,
their average clustering coefficients tend to not to be so aligned.
We denote the clustering coefficient of random counterparts as $C_{\rm rand}$. 
For epic narratives, clustering coefficient for friendly and full networks 
can be  an order of magnitude greater than those of random graphs of the same size and this is not the case for hostile network. 
Bylyny social networks are no different and values are given in Table~2.
These observations may serve as an evidence of the structural balance hypothesis for epics social networks: 
if two characters are connected through friendly links with a third one, it is likely that the first two are positively linked.
Likewise, if two characters are connected through hostile links with a third, it is unlikely that they are also  hostile to each other.

\subsection{Degree assortativity}\label{IVf}
% % % % % % % % % % % % % % % % % % % % % % % % % % % % % % % % % % % % % % % % % % 

In real social networks, 
individual nodes tend to be connected to  vertices of similar degree; 
(``birds of a feather flock together'')~\cite{Newman2003}.
In fictional social networks 
(such as the Marvel universe \cite{gleiser2007become}), 
however, high-degree vertices are mostly connected 
to low-degree ones (``not everyone can be a superhero'').
Social networks of epic narratives tend to occupy an intermediate position. 
A measure for correlation between neighbouring nodes 
is the degree assortativity $r_k$ \cite{newman2002assortative,Newman2003}.
By definition, degree assortativity is bounded so that $-1 < r_k < 1$. 
Positive values characterise assortative networks whereby high-degree 
nodes connect to each other 
and nodes with low degrees are interconnected as well.
Negative values, on the other hand, indicate that the network is disassortative, 
i.e.  that high-degree nodes are joined to nodes with a low degree and vice versa. 

Assortativity values for bylyny social networks are listed in Table \ref{tab2}, 
along with those of other epics. 
For the full network, a value  $r_k=-0.15$  indicates that the network is disassortative. 
This result is consistent with the fact that bylyny are epic tales about heroes: 
parts of networks are star-shaped structures (Figure~\ref{fig1}) 
where heroes associate with less significant characters. 
A  different behaviour is found for some Icelandic sagas;
they are stories about people and for them the assortativity is positive.
In a previous network study of the Icelandic sagas we concluded that network analysis allows to quantitatively compare societies from epic narratives with a real-world social networks~\cite{MacCarron2013}.

In epic networks, the hero is sometimes brought into contact with lower-degree characters through conflict.
This results in a greater degree of negativity in the degree assortativity of the hostile networks than the friendly one. 
This tendency is evident in Beowulf and the Iliad. 
It does not occur in bylyny, however -- or in Ireland's T{\'{a}}in B{\'{o}} C{\'{u}}ailnge.
In this sense, again,  bylyny find similarities with Ireland's iconic tale.
% The results listed in Table \ref{tab2}, then, with the exception of Icelandic sagas, indicate that most social networks of epic narratives are disassortative. 
It should be noted, however, that the negativity of assortativity 
can sometimes be caused by the structure of the story itself. 
For example, events described in the epic Beowulf occur in two different 
places and are also divided by time intervals. 
They are connected only through a common protagonist. 
As shown in Refs.\cite{MacCarron2012,Kenna2016}, 
removal of the protagonist changes the nature of 
network connections from disassortative to (weakly) assortative. 
A similar effect is present in the bylyny social network. 
Removal of the node corresponding to Prince Volodymyr increases the  degree assortativity to $r_k=0.01$, 
which is indeed a positive value, albeit a very small one. 

As discussed in Section 2, many researchers believe that 
Prince Volodymyr is a collective character \cite{Shevchuk2003} --- 
an amalgamation of historical figures Volodymyr the Great, Yaroslav the Wise, and Volodymyr Monomakh \cite{IEU}}. 
We split Prince Volodymyr's node into three distinct vertices and randomly distributed his edges amongst them. 
We did this multiple times and find that the bylyny network assortativity increases assortativity to $r_k=0.004$. 
In this way, the network becomes more similar to a real-world social networks. 
Division into four makes the network even more assortative with $r_k=0.03$. 
Thus, we see that dividing the superheroic main character into more ``human'' (less connected) individuals  makes the network more akin to a real social network.
Of course, disassortativity is necessarily driven by high-degree characters  interacting with low degree ones and reducing the degree of the most connected characters necessarily increases the value of $r_k$. 
In a sense this procedure  ``de-mythicises'' the hero, so it is up to the reader to make what they will of it.
Certainly, and as always, we  advise that any conclusion cannot be drawn from a single metric and has to involve holistic input  from humanities.

\subsection{Communities}\label{IVi}

% % % % % % % % % % % % % % % % % % % % % % % % % % % % % % % % % % % % % % % % % % 

\begin{figure}[ht]
 \centerline{\includegraphics[width=12.0cm]{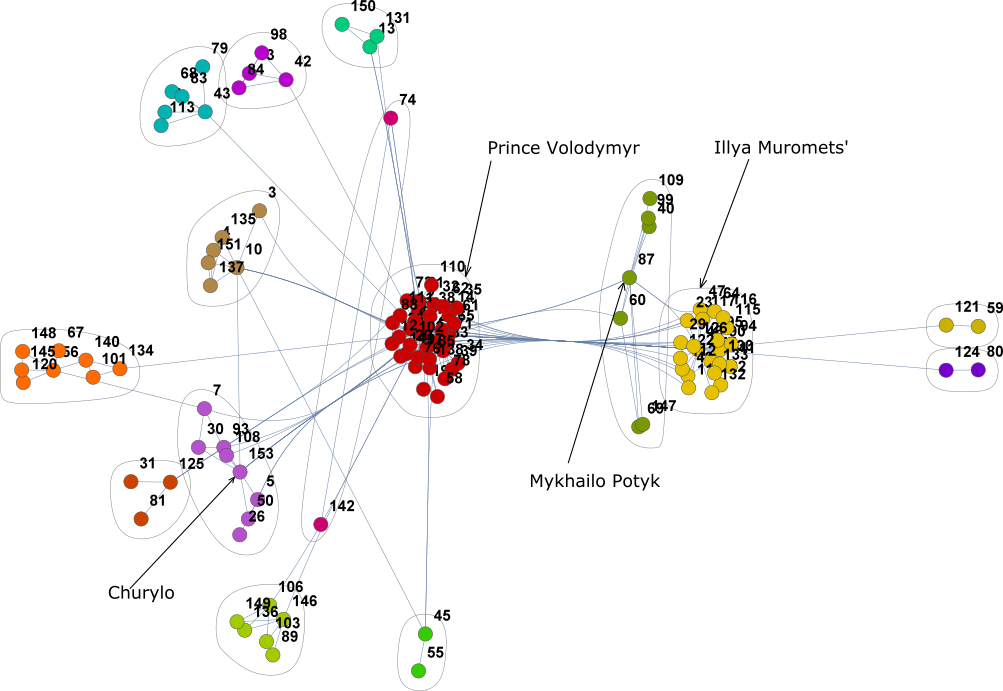}}
  \caption{Community structure of the bylyny social network 
  (with both hostile and friendly links).
  The largest of 15 communities is centered around Prince Volodymyr 
  (\#52 in the center of the picture), 
  while that around Illya Muromets' is second biggest (\#46 on the right side of the figure).  
  The next two communities by size are concentrated around Churylo 
  (\#153) and Mykhailo Potyk (\#87). }
 \label{fig4}
\end{figure}

Even from  Figure~\ref{fig1} it is obvious that the bylyny network comprises distinct communities. 
While this visualisation depicts interacting characters,  it hides a more subtle  community structure --- for, even in the giant component many vertices are more heavily connected certain others than with the rest of the network. 
The standard approach to  detecting communities uses a specially tailored algorithm created by Girvan and Newman \cite{newman2004finding}. 
Application to the bylyny social network (taking into account both types of links) yields an interesting result: 
the largest connected components is divided into 15 communities. 
We depict these in Figure~\ref{fig4}.
The largest such community has 36 nodes (coloured red in the center of the figure).
These include Prince Volodymyr (\#52), his wife {Opraksiya} (\#104), Oleksiy Popovych (\#1) and Dobrynya (\#28). 
The second largest community has 22 characters.
This is the large group of yellow nodes concentrated around Illya Muromets' (\#46) to the right side of the figure. 
This demonstrates that 
Illya Muromets' (\#46) and Dobrynya {(\#28)}
with Oleksiy Popovych (\#1) 
represent a different generations of heroes \cite{Putilov1986,Avenarius1902}. 
Illya is from the older heroes,  whereas Oleksiy and Dobrynya are from the  younger ones.

The analysis also identifies a community concentrated around Churylo (\#153) and another around Mykhailo Potyk (\#87). 
Both of these characters, along with Duke (\#33) {are associated with} Halicia–Volhynia \cite{Grushevskyi}. 
This  Principality or Kingdom was a medieval state and vassal of the Golden Horde that existed from 1199 to 1349.
The appearance of these characters in the epic describing the Kyivan Rus', 
can be explained by the fact that after the destruction of Kyiv by the Mongols, 
cultural heritage passed to  Western 
land in the Halicia–Volhynia kingdom and it has absorbed part of the local heritage. 
As one can see, the quantitative analysis of bylyny social 
network also provides separate communities centred around these characters, as shown in 
Figure~\ref{fig4}.

% % % % % % % % % % % % % % % % % % % % % % % % % % % % % % % % % % % % % % % % % % 
\section{Conclusions and Outlook for the Future}\label{V}
% % % % % % % % % % % % % % % % % % % % % % % % % % % % % % % % % % % % % % % % % % 

To what extent do epic narratives belonging to different cultures share common and differentiating characteristics, 
and to what extent can we quantify them?
This was the question asked 10 years ago in Ref.\cite{MacCarron2012}.
In the interim years, the approach has been developed, extended, 
improved and applied to an increasing corpus of (mostly medieval) literature \cite{MacCarron2012,MacCarron2013,MacCarron2014,Kenna2016,Yose2016}. 
This paper brings the approach to bylyny -- the heroic epics of Eastern Slavs. 

Table~\ref{tab2} captures some network properties of the societies in this and other epic and mythological narratives. 
All full and friendly networks are small world in that average path lengths are similar to random average path length and they have  high clustering coefficient.
Additionally, their maximum degrees are much greater than the average, indicating the degree distributions all have right skew. 
These are unifying characteristics - placing bylyny firmly in the category of other European and world narratives.

It is the assortativity that shows the key differences between these networks - but even these differences unite.
As discussed in Ref.~\cite{MacCarron2014},  the social networks properties often reflect the type of story. 
For example, Iceland's Nj\'als saga and Laxd{\ae}la Saga are stories mostly of a people and set in a single location. 
Beowulf and the Odyssey, for example, are focused on a protagonist who travels between locations. 
Narratives like the Iliad and  Egils saga lie between these two in that they have protagonists with relatively large degrees 
(decreasing the value of $r_k$) and their friendly networks are assortative. 
Bylyny are mostly focused on a single node making it more similar to these heroic epics than the other two types of story. 
It also shares many network properties with Ireland's T{\'{a}}in B{\'{o}} C{\'{u}}ailnge and the Ulster cycle of Irish mythology.

Vasyl Avenarius in the introduction to his book \cite{Avenarius1902} notes that Prince Volodymyr is called ``the sun'' because all 
the characters orbit around him like the planets orbit around the sun. 
Our results support this opinion. 
Prince Volodymyr is the most important node by degree, closeness and betweenness centralities.
Bylyny social networks are disassortative by nodes degree, a property that is shared with most of the epics in Table~\ref{tab1}.
%Real social networks, on the other hand, tend to be assortative \cite{newman2002assortative,Newman2003}). https://www.overleaf.com/project/61c20732ff8fd812117aab88
As with the character Beowulf in the eponymous tale, by removing him we see the rest of the social network is assortative~\cite{MacCarron2012}, 
Prince Volodymyr therefore drives the disassortativity and  the  previous suggestion that he is an amalgamation (a suggestion similar to one made previously for the Irish T{\'{a}}in B{\'{o}} Cuailnge~\cite{MacCarron2012}),  reduces his degree and therefore the assortativity, making the network more like that of the Iliad and Egils saga.
Indeed, the friendly network with Prince Volodymyr node removed, becomes slightly assortative. 

Thus the Kyiv bylyny cycle of east Slavic epic narratives falls nicely within the European tradition in network terms --- it is in many ways like Ireland's heroic tradition and Iceland's social ones.

~\\

{\bf{Author Contributions}}:
Data were collected and translations from chronicles 
%in Appendix X
were performed by Petro Sarkanych.
Most of the
humanities contextual information was provided by Nazar Fedorak.
The remaining authors contributed to the technical side, interpretation and writing of the paper.
Although a number of years in germination, this paper was completed in haste as half its authors were in a warzone.
Thus we were limited in the extent to which we could discuss the final version.
The last author takes full responsibility for any errors and omissions.
All suggestions for improvements and corrections should be addressed to him.
%A follow-on paper will appear in due course.

~\\

{\bf{Acknowledgements}}:
We would like to thank 
Bertrand Berche,
Reinhard Folk,
Julian Honchar, 
Maddy Janickyj,
Taras Yavorskyi
for academic  discussions. The project is part supported by Coventry University's allocation of UKRI Participatory Research funding. Yu.H. acknowledges support
of the JESH mobility program of the Austrian Academy of Sciences and hospitality
of the Complexity Science Hub Vienna when finalizing this paper.
We thank people of Ukraine for inspirational bylyny and for inspirational heroism and defence of freedom, including academic freedom.
We thank many academics from other countries --- including some Slavic ones ---
%in other Slavic countries
who have sent public and private messages of support, despite some being  neither free nor academically free to speak out.
We look forward to collaborating with them in future
enlightening all of our populations so that mistakes of the past are left there.

~\\

{\bf{Special Note}}:
National epics like the ones discussed in this paper are important elements of national identity. Many countries use representations of iconic characters from the mythological or historical past in identity documents, iconography and coinage. In his recent essay, ``On the Historical Unity of Russians and Ukrainians'', Vladimir Putin invoked the distant past when he said~\cite{Putin} 
``The spiritual choice made by St. Vladimir, who was both Prince of Novgorod and Grand Prince of Kiev, still largely determines our affinity today.''\footnote{The Kyivan Rus' name ``Volodymyr'' transform to ``Vladimir'' in the Russian language. Likewise ``Kiev'' is the Russian spelling for ``Kyiv''.} 
This is a basis for his view that 
``the idea of Ukrainian people as a nation separate from the Russians [has] no historical basis – and could not have been any.'' 
Putin's erroneous view of culture and history has been roundly refuted in multiple authoritative quarters. 
In his subsequent interview,  author and historian Timothy Snyder says of Putin~\cite{Polish}: 
``He considers himself the second Volodymyr the Great, and sees his task as completing his work, which began more than a thousand years ago.'' 

However, as we see in this paper, Prince Volodymyr is a  character strongly connected with people of different opinions --- people such as  Illya, ``who does not always blindly perform the orders of  prince Volodymyr and sometimes transforms them or even resists them.'' 
%In the future, the narrative of the current political
%situation will be written by historians and other academics. 
%As for the heroic Kyiv
%cycle of Ukrainian bylyny, we expect that heroes will again embrace freedom of
%thought and dissenting opinions; and will hail from a free and peaceful Kyiv.
In the future, the narrative of the current political situation will be written by historians and other academics, and not by autocrats or sycophants.
As for the heroic Kyiv cycle of Ukrainian bylyny, 
we expect that that story will be of heroes 
again embracing freedom of thought and 
dissenting opinions; and will centered on a free and peaceful Kyiv.

\bibliographystyle{ws-acs}
\bibliography{acs3}

\section*{Appendix A: Dobrynya in epics and in chronicles}

%\red{RK asks: What should be the title of this apendix? } %\blue{YuH suggests: Dobrynya in epics and in chronicles. PS: agree with this one}
%\blue{PS: I believe, this was to add context to the points, where we discuss possible connections of bylyny characters with the historical figures.}

In humanities literature, 
the correspondence between the epic Dobrynya and the chronicle Dobrynya
(from the Tale of Bygone Years) is frequently studied, in attempts to determine
the extent to which the former is drawn from the latter.
In the chronicles, Dobrynya is a brother of Malusha 
(the wife of Prince Svyatoslav, who himself is the father of Volodymyr the Great).
%, therefore, the uncle of Volodymyr-Vasyl the Great. 
He was a member of the old aristocracy (a “boyar”), 
a warlord (“vojevoda”) of Kyiv and a representative (“posadnyk”) of Volodymyr  
the Great in Novgorod.
%The first mention of Dobrynya is dated in the chronicle as the year 970:
%\red{according to the chronicles,}
``In the year 6478 [970] Svyatoslav sent Yaropolk to Kyiv, 
and Oleh to Derevlyans as his princes.
At the same time, people of Novgorod came, asking the prince for themselves: 
`If you do not go to us, then we will find the prince by ourselves.' 
Svyatoslav replied that he will find them a prince. 
Yaropolk and Oleh refused, so Dobrynya suggested: `You should ask Volodymyr' - because Volodymyr was son of Malusha, sister of Dobrynya. 
Therefore Dobrynya was an uncle of Volodymyr. 
The people of Novgorod were happy with Svyatoslav sending Volodymyr to them.
In the end, Volodymyr was sent to Novgorod together with his uncle Dobrynya, 
and Svyatoslav was sent to Pereyaslavets' '' \cite{Machnovec1989}.

Ten years later, 
in the year  980 according to the chronicles, 
Volodymyr started a campaign in Novgorod, 
where, at that time, Prince Yaropolk  already was ruling. 
Here is the beginning of this story [the translation is ours]:
\begin{quote}
``Volodymyr came with the Varangians [Vikings] to Novgorod and said to Yaropolk’s prince:
  `Go and tell my brother: 
     ``Volodymyr is going against you, prepare to fight.'''
He then sat in Novgorod, and sent a [young] messenger 
to Rogvolod the prince of Polotsk, saying: 
  `I want to take your daughter Rognida as a wife.' 
Rogvolod then asked his daughter: 
  `Do you want to merry Volodymyr?'
She replied:
  `I do not want to take off Volodymyrs shoes, but Yaropolk's.'
  %\footnote{This means that Rognida didn’t want to become Volodymyrs wife, but preferred to merry his brother Yaropolk.} 
Rogvold came from across the sea and had his volost'
in Polotsk.\footnote{A Volost' was a traditional administrative subdivision in the Kyivan Rus’, usually ruled by a local prince.}   ...

The messenger came back to Volodymyr, 
and told him the full  answer of Rognida, 
the daughter of Rogvolod the Prince of Polotsk. 
Volodymyr gathered a huge army including Varangians, 
Slavs, Chood' and Kryvychi - and went agains Rogvolod. 
At the same time Rognida was preparing to be sent to meet Yaropolk. 
But Volodymyr came to Polotsk, and killed Rogvolod and his two sons. 
He also took Rognida for the wife and continued his campaign against Yaropolk''.\cite{Machnovec1989}
    \end{quote}

In the same year, according to the Tale of Bygone Years, 
Volodymyr, 
having defeated Yaropolk and finally becoming established in Novgorod, 
sat on the princely throne in Kyiv.
\begin{quote}  ``Volodymyr began to reign in Kyiv alone. 
   He set his idols on a hill outside the palace courtyard: 
    the Perun\footnote{Here and below follow names of principal pagan deities of the Slavs: Perun, Khors, Dazhboh, Striboh, Simargal, and Mokosh.}  
    of the wood, and the head was silver, and the moustache of gold; 
    Hors, and Dazhbog, and Strybog, and Simargal, and Mokosh. 
    People were bringing human sacrifices to these idols, 
    calling them gods, but sacrificing to demons. 
    The land of the Rus' and the hill were defiled by their sacrifices. 
    But the precious god does not want death of sinners; 
    on this hill there is now the Church of St. Vasyl the Great\footnote{Named after Prince Volodymyr second name Vasyl he got after baptisation.} 
   ... ''~\cite{Machnovec1989}.
\end{quote}
Meanwhile, 
  \begin{quote}
   ``Volodymyr has sent Dobrynya, his uncle, to Novgorod.
   Dobrynya, having come to Novgorod, placed the idol of Perun on the Volkhov River, 
   and brought sacrifices of the Novgorod people as to this idol as God ''~\cite{Machnovec1989}. 
   \end{quote}
Here we see ``uncle Dobrynya'' as the most trusted person of Grand Prince Volodymyr;
 he is trusted both politically and religiously - even in a pagan 
 %co-ordinates.
 sphere.
 
The next mention of Dobrynya comes five years later:
  \begin{quote}
  "In the year 6493 [985] Volodymyr started a campaign to Bulgaria with Dobrynya, 
   his uncle, in boats, while Torks  travelled by horses. 
   After defeating the Bulgarians, Dobrynya said to Volodymyr: 
      `I looked at the prisoners, and they all were in boots. 
      We will not take these as a tribute, we both go to look for those in the carbatinas.' ''\cite{Machnovec1989}
 \end{quote}
 This mentioning is interesting to us because Dobrynya is not only a close proxy, 
 but also  Volodymyr’s counsellor, 
 endowed with observation and ability to make non-obvious conclusions.

The chronicle does not cover the death of  Dobrynya. 
Together with the death of Grand Prince Volodymyr-Vasyl the Great in the year 1015, his uncle also disappears from the pages of the Tale of Bygone Years. 
But already in 1018 we read about Dobrynya's son named Constantin. 
Here is this episode where the sons of Prince Volodymyr (Yaroslav, later named ``the Wise'') and Dobrynya met each other: 
\begin{quote}
  ``When Yaroslav came to Novgorod, he wanted to flee the sea. 
    Local prince Constantin, son of Dobrynya, 
    with the Novgorodians destroyed the boats and said to Yaroslav: 
    `We can still fight for you with Boleslav and Svyatopolk.' 
    Novgorodians began to collect money: from a man - four coins;  from an elder - ten hryvnias;  and from a boyar - eighty hryvnias. 
  This allowed Yaroslav to hire Varganians and to gather big army."\cite{Machnovec1989}
\end{quote}
Dobrynya son's Christian name draws attention here --- it indicates that Dobrynya himself was baptized with Volodymyr in 988. 
Afterwards he gave his son not a pagan Slavic name but a Christian one.

For completeness of the picture, we should 
mention another character with the name Dobrynya from the ``Tale of Bygone Years''.
This is Dobrynya Raguilovych another prince of Novgorod, 
but in the year of 1096. 
\begin{quote}
 ``Oleh sent his brother Yaroslav [Svyatoslavych] to the vanguard. 
  He stood on the field near Rostov. 
  Mstyslav in the meantime consulted with Novgorodians. 
  They sent in front of him Dobrynya Raguilovych, 
  and Dobrynya first of all captured the collectors of the tribute. 
  When Yaroslav heard that collectors were captured, 
  because Yaroslav stood at the guard on the river, 
  he ran away that night and ran to Oleh and told him that Mstyslav was coming, 
  There was also a message to Oleh that his watchmen were captured, 
  so he moved to Rostov.''~\cite{Machnovec1989}.    
\end{quote} 
Not going deep into the historical details of this message, it is worth noting that Dobrynya Raguilovych mentioned here 
is also obviously a brave and smart man 
(judging by the description of his deeds) 
and that he was a contemporary of another well-known Prince Volodymyr – Volodymyr Monomakh.  
So different Dobrynyas might have followed the same path as different Volodymyrs and became a single character.
We will endeavour to investigate these possibilities in a further study.

%%%%%%%%%%%%%%%%%%%%%%%%%%%%%%%%%%%%%%%%%%%%%%%%%%%%%%%%%%%%%55
\section*{Appendix B: Bylyny giant component}
%with weighted edges
%%%%%%%%%%%%%%%%%%%%%%%%%%%%%%%%%%%%%%%%%%%%%%%%%%%%%%%%%%%%%55

\begin{figure}[!ht]
  \centerline{\includegraphics[width=12.0cm]{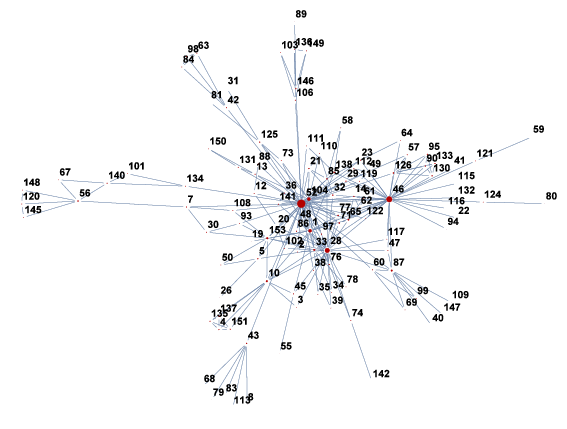}}
%  \includegraphics[width=6.0cm]{2b.png}}
%  \centerline{(a) \hspace{17em} (b)}
%  \centerline{\includegraphics[width=6.0cm]{2c.png}
%\includegraphics[width=6.0cm]{2d.png}}
%  \centerline{(c) \hspace{17em} (d)}
  \caption{The largest connected component of the bylyny network   with vertices sized proportionate to node degree $k$.}
  \label{figA1}
 \end{figure}

\begin{figure}[!ht]
  \centerline{%\includegraphics[width=12.0cm]{2a.png}}
%  \includegraphics[width=6.0cm]{2b.png}}
%  \centerline{(a) \hspace{17em} (b)}
%  \centerline{\includegraphics[width=6.0cm]{2c.png}
 \includegraphics[width=12.0cm]{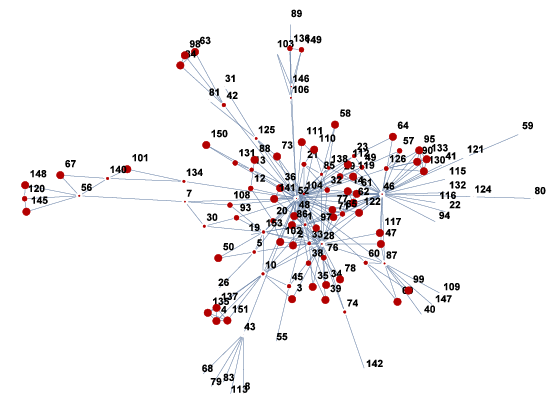}}
%  \centerline{(c) \hspace{17em} (d)}
  \caption{The giant component with nodes weighted by clustering coefficient $C$.
  }
  \label{figA2}
 \end{figure}

In Figure~\ref{figA1} we present 
the largest connected component
of the bylyny network 
(including both friendly and hostile links) 
with vertices sized proportionate to 
node degree $k$.
%(a); closeness centrality $C_C$ (b); betweenness centrality $C_B$ (c) and clustering coefficient $C$ (d). 
% Different characters rank differently according to different measures. 
Prince Volodymyr (\#52) and Illya Muromets' (\#46) have the highest degrees and corresponding nodes are largest.
They also have highest  closeness and betweenness  centralities  ($C_C$ and $C_B$, respectively) and similar plots for these  measures can be constructed.
Their clustering coefficients are small, however and ac ounterpart plot for this measure is given in Figure~\ref{figA2}.
 Likewise Ivan Hostynyj Syn (\#43) has a relatively high centrality values but low node degree and clustering coefficient.
 %Such plots are available on request.
% \footnote{\blue{PS said: added file 2aresc.png with rescaled node sizes. \red{RK can't find that file! }}}

\begin{figure}[!ht]
  \centerline{\includegraphics[width=8.0cm]{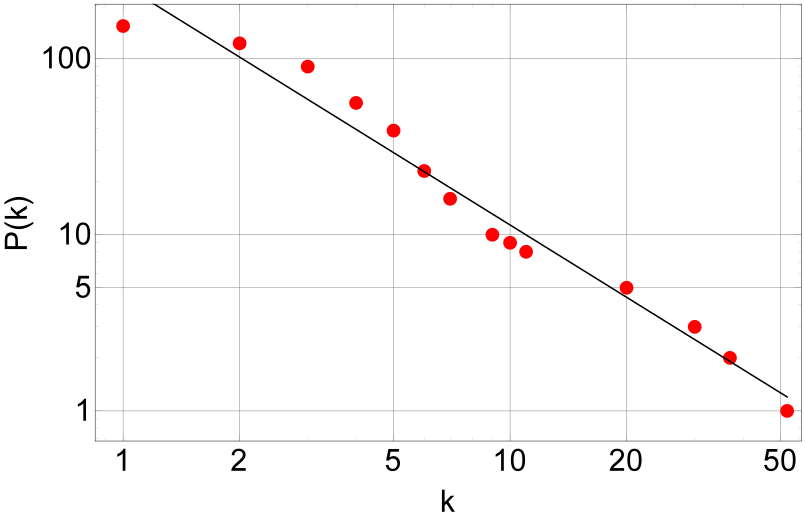}}
%  \includegraphics[width=6.0cm]{2b.png}}
%  \centerline{(a) \hspace{17em} (b)}
%  \centerline{\includegraphics[width=6.0cm]{2c.png}\includegraphics[width=6.0cm]{2d.png}}
%  \centerline{(c) \hspace{17em} (d)}
  \caption{Cumulative degree distribution $P(k)$ as a function of a node degree in a double logarithmic scale. Straight black line shows a fit to a power-law suggesting a scale-free property of the network.}
  \label{17Mar}
 \end{figure}

 Finally, in Figure~\ref{17Mar}, we give a plot of the cumulative degree distribution plotted against node degree. 
 The fit is a power-law, suggestive of a scale-free properties of the bylyny character network.
 We refer the reader to the literature for similar plots for other epic tales \cite{kenna2017networks}.

\end{document}